\documentclass[twocolumn,
preprintnumbers,
 amsmath,amssymb,
 aps,bm
]{revtex4-2}

\usepackage{amsfonts}
\usepackage[caption = false]{subfig}
\usepackage{graphicx}
\usepackage{dcolumn}
\usepackage{bm}
\usepackage{hyperref}
\usepackage{cancel}

\def\ltap{\ \raise.3ex\hbox{$<$\kern-.75em\lower1ex\hbox{$\sim$}}\ }
\def\gtap{\ \raise.3ex\hbox{$>$\kern-.75em\lower1ex\hbox{$\sim$}}\ }
\def\lsim{\ \raise.3ex\hbox{$<$\kern-.75em\lower1ex\hbox{$\sim$}}\ }
\def\gsim{\ \raise.3ex\hbox{$>$\kern-.75em\lower1ex\hbox{$\sim$}}\ }
\newcommand{\eg}{{\it e.g.}}
\newcommand{\ie}{{\it i.e.}}

\newcommand{\be}{\begin{equation}}
\newcommand{\ee}{\end{equation}}
\newcommand{\beq}{\begin{equation}}
\newcommand{\eeq}{\end{equation}}
\newcommand{\bea}{\begin{eqnarray}}
\newcommand{\eea}{\end{eqnarray}}
\newcommand{\bear}{\begin{eqnarray}}
\newcommand{\eear}{\end{eqnarray}}
\newcommand{\mstar}{\ensuremath{M_*}}
\newcommand{\rstar}{\ensuremath{R_*}}

\newcommand{\eref}[1]{(\ref{#1})}

\def\keV{\,{\rm keV}}
\def\MeV{\,{\rm MeV}}

\def\mnras{MNRAS}

\newcommand{\Msun}{\ensuremath{M_\odot}}

\def\MeV{\,{\rm MeV}}

\def\mnras{MNRAS}

\usepackage[usenames,dvipsnames]{xcolor}

\begin{document}

\preprint{FERMILAB-PUB-23-029-T}

\title{Recurrent Axinovae and their Cosmological Constraints}

\author{Patrick J. Fox$^1$}
\author{Neal Weiner$^2$}
\author{Huangyu Xiao$^3$}
 \affiliation{$^1$Particle Theory Department, Theory Division, Fermilab, Batavia, IL 60510, USA}
\affiliation{$^2$Center for Cosmology and Particle Physics, Department of Physics, New York University, New York, NY 10003, USA}
\affiliation{$^3$ Astrophysics Theory Department, Theory Division, Fermilab, Batavia, IL 60510, USA}

\date{\today}

\begin{abstract}
Axion-like dark matter whose symmetry breaking occurs after the end of inflation predicts enhanced primordial density fluctuations at small scales.  This leads to dense axion minihalos (or miniclusters) forming early in the history of the Universe. 
Condensation of axions in the minihalos leads to the formation and subsequent growth of axion stars at the cores of these halos.  If, like the QCD axion, the axion-like particle has attractive self-interactions there is a maximal mass for these stars, above which the star rapidly shrinks and converts an $\mathcal{O}(1)$ fraction of its mass into unbound relativistic axions.  This process would leave a similar (although in principle distinct) signature in cosmological observables as a decaying dark matter fraction, and thus is strongly constrained.  We place new limits on the properties of axion-like particles that are independent of their non-gravitational couplings to the standard model.
\end{abstract}

\maketitle

\section{Introduction}
The stars in our galaxy are made of $\sim 10^{57}$ fermions bound together by gravity and protected from collapse by thermal pressure or fermion degeneracy pressure.  In the presence of a light, long-lived boson similar gravitationally bound states of that boson may exist, but in the absence of nuclear burning they are instead supported by gradient pressure, which is a result of the uncertainty principle.  Axion stars are one such example of these bosonic objects. 

In the Standard Model, stars convert approximately $0.1\%$ of their mass energy into radiation over their lifetime. The small energy released
(compared to rest mass) in the pp-chain, for instance, is due to the relatively small binding energy inside the star. Only very compact objects like neutron stars are relativistic in nature. Moreover, because of baryon number conservation, there is a limitation on overall energy release given the (approximately degenerate) neutrons and protons which must remain in the final state.

However, in the dark sector, there are reasons to expect the overall energy conversion could be much higher if a similar process were to occur. Complete conversion of rest mass from \eg\ a $3\rightarrow1$ process is possible because there is no ``baryon-number" conservation for bosonic dark matter. For example, axion stars will collapse and emit relativistic axions when they reach a critical mass. We call such processes that drastically convert dark matter to dark radiation as \textit{Axinovae}. There is no mechanism to quench the axionovae if axion stars form ubiquitously in the Universe, as expected in the post-inflationary scenario where axion miniclusters form after matter-radiation equality. Therefore, a large formation rate of axion stars that lead to axinovae is very constraining. We take the formation of axion stars as a concrete example to study but the result can apply to generic scalars whose self-interaction is attractive since the properties of axion stars do not depend on any interactions other than gravity and the axion self-couplings.

A natural cosmic history that can occur generically for these models is, after matter-radiation equality, these axions stars form, grow, and finally explode as an axinova, converting a significant fraction of energy into semi-relativistic axions. After this, the remnant can continue to grow, until it explodes again. This process of recurrent axinovae can convert a significant fraction of the dark matter into relativistic energy, which is then constrained by cosmological observations.

This paper is organized as follows: in Sec.~\ref{sec:axions}, we discuss the formation of enhanced structures at small scales due to the axion perturbations and study the formation history of axion stars inside those structures. In Sec.~\ref{sec:constraints}, we study the constraints on axion parameter space by requiring the decay fraction of axion dark matter should not exceed an upper bound. In Sec.~\ref{sec:conclusions}, we present our conclusions.

\section{Axions, Axion minihalos, and axion stars}\label{sec:axions}
The axion is a well-motivated dark matter candidate, which can also leave unique fingerprint on the matter power spectrum at small scales if the PQ symmetry breaking occurs \emph{after} inflation. In such scenarios, different horizon patches have different matter densities when the axion acquires its mass, leading to the formation of axion miniclusters or axion minihalos at matter-radiation equality \cite{Hogan:1988mp,Kolb:1995bu}. More interestingly, coherent objects called axion stars can form in the center of axion minihalos due to Bose-Einstein condensation \cite{Kolb:1993zz}, which may eventually accrete into a critical object and emit relativistic axions. We call such phenomenon axinovae, which can occur with an attractive axion self-coupling and the formation of axion minihalos at matter-radiation equality.

Originally proposed to solve the strong CP problem \cite{Peccei:1977hh,Weinberg:1977ma,Wilczek:1977pj}, the present-day landscape of axions and axion-like particles (ALPs) is broad.  One common feature across this landscape is that the axion, $\phi$, is a pseudo-Goldstone boson of a global $U(1)_{PQ}$ symmetry broken at a scale $f_a$.  The $U(1)_{PQ}$ is anomalous under a confining gauge group which means that the axion's potential is generated through instanton effects occurring at the compositeness scale of the gauge group, $\Lambda$, and takes the form
\be
\label{eq:VaxionFull}
V(\phi) =\frac{\Lambda^4}{c_{ud}}\sqrt{1-4c_{ud}\, \sin^2\left(\frac{\phi}{f_a}\right)}~.
\ee
In the case of the QCD axion $\Lambda\approx 200\MeV$ and $c_{ud}\approx m_u m_d/(m_u+m_d)^2 \approx 0.2$.  In addition to the self couplings, coupling to gravity, and the anomaly-induced coupling to QCD (or QCD-like group), the axion may have model-dependent couplings to other SM gauge bosons and fermions.  We focus here on the self and gravitational couplings only, which can already lead to interesting dynamics such as axinovae.

\subsection{Axion Minihalos}
In the post-inflationary scenario, the present-day Universe contains a large number of patches which were causally disconnected at the time of QCD phase transition.  In each causally disconnected patch of the Universe, axion field values are uncorrelated.  Once the axion acquires a mass, and Hubble friction is small enough, the axion behaves as cold dark matter and isocurvature fluctuations are present in the matter density. When the Universe becomes matter dominated this small-scale structure will start to collapse under gravity, leading to axion minihalos.
Furthermore, there may be large overdensities of axions at even smaller scales arising from the evolution of the network of axion strings and domain walls \cite{Kibble:1976sj} set up when the PQ symmetry breaks.  Even for the much studied case of the QCD axion, there is controversy \cite{Gorghetto:2018myk,Klaer:2017ond,Fleury:2015aca,Chang:1998tb,Hagmann:2000ja,Buschmann:2021sdq,Vaquero:2018tib,Buschmann:2019icd,Gorghetto:2020qws,Kawasaki:2018bzv,Hiramatsu:2010yu,Fleury:2015aca,Kawasaki:2018bzv,Klaer:2019fxc,Vaquero:2018tib,Buschmann:2019icd,Gorghetto:2020qws,Buschmann:2021sdq,Hindmarsh:2019csc,Hindmarsh:2021vih,Hindmarsh:2021zkt} as to what fraction of the relic dark matter axions arise from misalignment or from the decay of topological defects. Along with those topological defects, objects called oscillons or axitons that can contribute to the small scale overdensities will form after the axion accquires its mass \cite{Buschmann:2019icd,Vaquero:2018tib}. Those objects can form when the axion self-interaction dominates over the Hubble expansion term, which is easily satisfied in the early Universe when the self-interaction is strong due to the high density. As the axion density drops, the formation of oscillons will be turned off and oscillons themselves will dissipate via emitting relativistic axions. 

It is worth noting that the post-inflationary scenario is not essential for the axinovae. Any  matter power spectrum which is enhanced at small scales can lead to the formation of axion minihalos around matter-radiation equality, but the post-inflationary scenario is a minimal realisation.
We take a simple ansatz for the spectrum of initial fluctuations in the axion field, namely that the spectrum of isocurvature fluctuations in the axion field follow a white-noise spectrum, cut off at small scales \ie
\be
\label{eq:isowhitenoisepower}
\frac{\delta \rho_a}{\rho_a} = A_0\left(\frac{k}{k_0}\right)^3\, \Theta(k_0-k) ~.
\ee
Here $k_0 \approx a_{\mathrm{osc}} H_{\mathrm{osc}}$is the (comoving) wavenumber determined by the horizon size at the time the axion starts to oscillate, \ie~ $m_a(T_{\mathrm{osc}})\sim 3H_{\mathrm{osc}}$.  While here we consider a pure white noise spectrum we extend this analysis to a more general power law spectrum in Appendix~\ref{app:PS_analysis}.
In reality one would expect a softening of the cutoff in the white noise power spectrum at small scales.  The exact details of how this occurs is related to the dynamics of string network and axitons, and is unknown.  It will not affect our conclusions, see Appendix~\ref{app:PS_analysis} for details.
As mentioned above, the contribution of strings and domain walls to the abundance of non-relativistic axions is uncertain and will impact the size of the power spectrum.  Simulations typically show the density perturbations have $A_0\sim 0.1$ at $k=a_{\mathrm{osc}} H_{\mathrm{osc}}$, but they also show larger sub-horizon (larger $k$) fluctuations.  These sub-horizon fluctuations can collapse earlier than those at the horizon scale, leading to high concentration minihalos.  These halos are at smaller scales, $k\ge a_{\mathrm{osc}} H_{\mathrm{osc}}$, and have larger $\delta \rho_a/\rho_a$ and lower axion speeds in the mini-halos, resulting in a faster star growth rate.  We take the conservative limit of holding $k_0=H_{\mathrm{osc}}$ as the scale at which $A_0=1$.
With this white noise power spectrum \eref{eq:isowhitenoisepower} the first structures, of mass $M_0$, form at redshift $z_c\approx A_0^{1/2}z_{\rm eq}$ and the characteristic structure mass, defined as the peak in the distribution $M dn/d\log M$, occurs at 
\be
\label{eq:Mpeak}
M_{peak}(z)= M_0\left(\frac{1+z_c}{1+z}\right)^2~.
\ee
%
Where 
\be
\label{eq:M0}
M_{0} = 6\pi^2\overline{\rho}_0\left(\frac{1}{k_0}\right)^3~,
\ee
is the co-moving mass in the horizon at the time the axion starts to roll and $\overline{\rho}_0$ is the present-day cosmological axion density.  
The minihalos have a distribution of masses but for simplicity we use the characteristic mass $M_{peak}$ to provide a measure of the overall behavior.

The growth continues till around $z\sim 10-20$ when the minihalos merge into standard CDM halos and their growth stalls \cite{Xiao:2021nkb}.  

We take the minihalos to have an NFW \cite{Navarro_1996} density profile, defined by a scale radius $r_s$ and density $\rho_s$, 
\be\label{eq:NFW}
\rho(r) = \frac{\rho_s}{\frac{r}{r_s}\left(1+\frac{r}{r_s}\right)^2}~.
\ee
At the scale radius the circular speed is given by 
\be\label{eq:circularspeed}
v_s^2 = 4\pi G_N \rho_s r_s^2\left(\log 4-1\right)~.
\ee

This speed will be relevant for the calculation of axion star formation rate, and in the minihalos that will form axion stars this speed is much smaller than typical speeds in the Milky Way.
Numerical studies have shown that the dark matter halos at the characteristic mass will first reach a concentration factor \footnote{The concentration is defined as the ratio $c=r_{200}/r_s$ where $r_{200}$ is the radius at which the halos average density is 200 times the background DM density.} $c\approx 4$ at the time of halo collapse \cite{2003MNRAS.339...12Z}. For lighter halos that grow more through accretion than mergers, the halo concentration will grow linearly with the scale factor due to the decreasing background density. In the next subsections we discuss further structures that can develop in the core of these minihalos. In addition to the low speeds in mini-halos, the scale density in the mini-halos that form early is large,
\be\label{eq:rhos}
\rho_s =\overline{\rho}_0 (1+z)^3 \frac{\Delta_{200}\, c^3}{3\left(\log(1+c)-\frac{c}{1+c}\right)}  ~,
\ee
where $\Delta_{200}\approx 200$ in the spherical collapse model. For mini-halos that collapse at $z_{\mathrm{eq}}$ this density is $\rho_s \approx 10^{14}\overline{\rho}_0$.

\subsection{Lifecycle of an Axion Star}
We now turn to the question of formation of axion stars at the core of the minihalos discussed above. The subsequent growth and explosion of axion stars (axinovae) will also be studied. There are a few timescales we will discuss that are relevant in the lifecycle of axion stars:
\begin{itemize}
	\item The condensation timescale from gravitational interactions
	\item The condensation timescale from axion self-interactions.
	\item The evaporation timescale of light axion stars.
	\item The Hubble time when the axion star formation is active.
\end{itemize}
We will discuss those timescales later in this subsection.
For the parameter space that axinovae can place meaningful constraints on, the axion self-interaction always dominates over gravity. 

As discussed in Appendix \ref{app:as_configuration}, there are two branches of axion star configurations: the dilute branch, which, below a certain mass, is stable and the dense branch which will explode and emit relativistic axions. There is a critical star mass \eref{eq:mstarmax} that separates the two branches, which we denote as $\mstar^{\mathrm max}$. Therefore,  if they continue to accrete mass, the lighter dilute axion stars will eventually become unstable in a minihalo environment.  Axion minihalos are ideal environments for the axion star formation because they are dense and cold, owing to the high collapse redshifts and small virial masses. When the star formation rate is sufficiently large, stars will form in the minihalo center and grow to a critical mass star if the minihalo is massive enough. The critical star will contract under self-interaction and gravity, converting a large fraction of its mass to relativistic axions. 
Until the axinovae consume most of the minihalo mass, axion minihalos remain ideal environments for the axion star formation and axinovae shall occur again within the same timescale.  Thus, we naturally expect the axinovae phenomenon to be recurrent, when the growth timescale is fast enough. 
The crucial calculation to determine the fate of axion stars is the formation rate in the minihalo center and the corresponding star mass.

Once minihalos exist, gravitational interactions or self-interactions can subsequently lead to the formation of Bose-Einstein condensed axion stars at their center.  The timescale for this formation, and subsequent growth, in an environment where the axions being captured have typical number density $n$ and speed $v$ is determined \cite{Levkov:2018kau,Eggemeier:2019jsu,Chen:2021oot,Kirkpatrick:2020fwd} by 
\be\label{eq:generaltimescale}
\tau  \sim \left(f_{\mathrm{BE}} n \sigma v\right)^{-1} ~.
\ee
With $\sigma$ the total scattering cross section.
This formation rate is Bose-enhanced from the naive expectation due to the large phase space density, $f_{\mathrm{BE}}=6\pi^2 n(m_a v)^{-3}$.  The gravitational Rutherford transport cross section is \mbox{$\sigma_{\mathrm{gr}}=8\pi (G_N m_a v^{-2})^2 \log(m_a v R)$}, where the Coulomb logarithm has been cutoff at a characteristic length scale of the minihalo, $R$.  Attractive self-couplings can also lead to formation and the scattering cross section is \mbox{$\sigma_{\mathrm{self}}=\lambda^2 m_a^{-2}/128\pi$}.  The total condensation time, considering both gravity and self-interaction, is
\begin{equation}
\tau=\frac{\tau_{\rm self}\tau_{gr}}{\tau_{\mathrm{gr}}+\tau_{\rm self}}.
\end{equation}
With each individual process having a timescale of 
\be
\label{eq:taugrav}
\tau_{\mathrm{gr}}   = \frac{b}{48\pi^3}\frac{m_a v^6}{G_N^2 n^2\log \left(m_a v R\right)}~,
\ee
for gravity, and 
\be\label{eq:tauself}
\tau_{\rm self}=\frac{64 d m_a^5 v^2}{3\pi n^2 \lambda^2}~.
\ee

The parameters $b,d\sim\mathcal{O}(1)$ are numerical coefficients that are extracted from numerical simulations \cite{Chen:2020cef}.  Comparing these two timescales, \eref{eq:taugrav} and \eref{eq:tauself}, we see that the self interactions will determine the axion star formation rate if $f_a \lsim M_{\mathrm{pl}} v$.  Furthermore, if the relevant speed is determined by gravitational collapse of a minihalo (\ref{eq:circularspeed}) then self interactions dominate in the limit $f_a^2 \lsim \rho_s r_s^2$.
When determining the gravitational relaxation timescale for formation of axion stars in minihalos we take, as typical, the densities and speeds at the scale radius, see Eqs.~(\ref{eq:NFW}) and (\ref{eq:circularspeed}).  

In addition to the timescale for axion star growth there is also a rate for evaporation of the star.  Axions in the halo that are not part of the star can collide with bound axions causing them to be ejected.  The rate for this process shrinks with axion star mass and is approximately \cite{Chan:2022bkz} $\Gamma_{\mathrm{evap}}\sim (m_a v \rstar)^2 \tau^{-1}$.  The competition between growth and evaporation means only axion stars above a certain mass will gain mass by gathering axions from the halo.  As observed in numerical simulations \cite{Levkov:2018kau,Chan:2022bkz} such stars first appear after time $\tau$ and then proceed to grow.  The growth is initially fast ($d\log M/dt$ is constant) but once the virial velocity of the minicluster falls below the speed of the axions in the axion star the rate of growth slows, $d\log M/dt$ becomes inversely proportional to (a power of) the star mass \cite{Chan:2022bkz} which results in the mass growing with time as a power law.  The characteristic axion star mass where this change in behavior occurs is obtained by equating the virial velocity of the minicluster with that of the axion star \cite{Levkov:2018kau,Chen:2020cef,Eggemeier:2019jsu,Chan:2022bkz,Arvanitaki:2019rax,Du:2023jxh} is
\be\label{eq:msat}
\overline{\mstar} \approx 3\rho_a^{1/6} G_N^{-1/2}m_a^{-1} M_h^{1/3}~,
\ee
where $M_h$ is the halo mass.  The behavior of the growth rate once the axions in the star are moving faster than those in the halo is not definitively known, and there is evidence that it may continue to evolve with star mass \cite{Chan:2022bkz}.  This would result in the mass growing as a power law with a running index.  However, to simplify our analysis and to partially account for the numerical uncertainties, we will use a single power law but consider a range of possible powers.  In particular, we parametrize the power-law mass growth as $\mstar =\overline{\mstar}\, (t/\tau)^{1/\alpha}$ and vary $\alpha$ in the range of 1 to 5.
With initial exponential growth followed by constant power law growth, the timescale to form an axion star at critical mass $\mstar^{\rm max}$ depends in which regime the critical mass falls.  Thus,
\be\label{eq:tcrit}
t_{\rm crit} = \tau \times \begin{cases}
\log\left(\overline{\mstar}/\mstar^{\rm max}\right) + 1,\, &  \quad \mstar^{\rm max} \le \overline{\mstar}\\
(\mstar^{\rm max}/\overline{\mstar})^\alpha,\, &  \quad \mstar^{\rm max} > \overline{\mstar}
\end{cases}. 
\ee

The numerical simulations discussed above have mostly been carried out assuming a homogenous gas of axions as the initial background upon which an axion star forms.  For stars that form in minihalos the gas has a density and velocity profile.  In Appendix~\ref{app:powerlawinhalos} we argue that for an NFW profile the exponential growth is replaced with a power law, and the whole growth becomes a single power law, with $\alpha =3/2$ when self interactions dominate. 
Given that the majority of the dark matter has collapsed into axion minihalos with a characteristic mass $M_{\mathrm{peak}}(z)$, the total fraction of dark matter rest mass that has been converted to kinetic energy per unit time can be calculated as 
\begin{equation}\label{eq:decayfrac}
\frac{d f_{\rm decay}}{dt}=\frac{\kappa\, \mstar^{\rm max}}{M_{\mathrm{peak}}(z)\, t_{\rm crit}}~,
\end{equation}
where $\kappa$ is the fraction of the axion star's mass that is converted to relativistic axions during axinovae.  From simulations of these processes \cite{Levkov:2016rkk}, it is seen that approximately $50\%$ of the star's mass is lost during the nova and of this about $20\%$ is in the form of relativistic axions, so $\kappa \approx 0.1$.  The time to reach a critical star given in (\ref{eq:tcrit}) assumes the star grows from an undistorted minihalo.  After the first axinova there is a remanent of mass $\sim 0.5 \mstar^{\rm max}$ already present and the time for this to grow to $\mstar^{\rm max}$ is slightly shorter than for the first star.  For the power law considered here this correction is small and we ignore it, assuming all subsequent stars take time $t_{\rm crit}$ to explode.


\section{Cosmological Constraints}\label{sec:constraints}
\subsection{The decay rate of axion stars}
The process of forming axion stars which subsequently become nova converts non-relativistic dark matter axions into boosted ($\gamma\sim \mathcal{O}(\mathrm{few}))$ axions. 
The kinetic energy of the outgoing axions will red-shift away after the scale factor has grown by $\sim \sqrt{\gamma}$ and thus the dark matter's contribution to the matter-energy budget is depleted.  Here we study the impact of the cumulative loss of mass in the dark sector but it is possible that the temporary existence of a new relativistic species may lead to a measurable effect on large-scale structure and is worthy of future study.


This process is closely related to the scenario of decaying dark matter, which is well constrained by recent cosmological data \cite{Poulin:2016nat,Bringmann:2018jpr,Nygaard:2020sow}.  For dark matter which decays after recombination, the decrease of the dark matter fraction will increase the angular diameter distance to the last scattering surface over time. Furthermore, the amount of CMB lensing is reduced due to a smaller gravitational potential than expected. This scenario is constrained by a combination of CMB \cite{Planck:2018vyg} and, for very long lived dark matter, SDSS \cite{BOSS:2012dmf} data.  If the decay of dark matter occurs well before recombination or even before matter-radiation equality, the primary effect of the decaying dark matter is to enhance $N_{\rm eff}$ since the decay products behave as dark radiation.  In the short-lived situation the constraints are primarily from CMB measurements.  We will be interested in the long-lived case, and in particular decays which occur after matter-radiation equality but are no longer ongoing.  The equivalent bound  \cite{Nygaard:2020sow} for decaying dark matter on the fraction of the initial amount of dark matter that can decay is 
\be\label{eq:DDMbound}
f_{dDM} \equiv \frac{\Omega_{dDM}}{\Omega_{dDM}+\Omega_{DM}}\le 2.62\% \quad(\mathrm{at}\  2 \sigma )~.
\ee
Although the cosmological evolution of the dark sectors for decaying dark matter and axinovae are not identical they are similar and since the above constraint is independent of decaying dark matter lifetime over a wide range of lifetimes we will use it to constrain axions.  We leave a more detailed 
numerical analysis, and an investigation of other possible signals, for future work.  Converting (\ref{eq:DDMbound}) to the case of axinova leads to the requirement that 
\be\label{eq:ourbound}
\int_{z_c}^{z=20} \!\!\!dz\, \frac{d f_{\rm decay}}{dz} \le 2.62\%~.
\ee
In the scenario of axinovae, the decay of dark matter occurs when axion miniclusters start to form, which is always after matter-radiation equality. 
To avoid the constraint of (\ref{eq:ourbound}) requires either that the formation rate of axion stars is too small to be cosmologically relevant or that the formed axion star mass is smaller than the critical mass so there are no axinovae.  Note that this bound does not rely upon there being a coupling to any SM particles \eg\ photons, gluons, or SM fermions. However, our constraints do rely on the assumption that axions do make up the dark matter relic abundance and that the fluctuations in the axion field are isocurvature in nature and approximately power law.  If axions make up a fraction of the dark matter, this can be encoded as a reduction of $\kappa$, see \eref{eq:decayfrac}, and a corresponding weakening of the bounds.  

For normal misalignment production of axions, where $\langle \theta^2\rangle \approx 4$ \cite{GrillidiCortona:2015jxo}, the typical initial halos that form have a mass that depends upon the horizon size when the axion starts to oscillate $m_a(T_{\mathrm{osc}})= 3 H(T_{\mathrm{osc}})/2$.  For the QCD axion, where the temperature dependence of the axion mass is known, this oscillation time is uniquely determined.  However, in more general axion scenarios the oscillation temperature, and therefore $M_0$, is a free parameter.   In the radiation dominated era $H(T)=\pi\left(8\pi g_*(T)/90\right)^{1/2}T^2/M_{Pl}$ and the halos form with mass,
\be
\label{eq:M0Tosc}
M_h  = \frac{4\pi}{3}\left(\frac{1}{a(T_{\mathrm{osc}})H(T_{\mathrm{osc}})}\right)^3\overline{\rho}_0 \approx 2\times 10^8 \Msun \left(\frac{\mathrm{keV}}{T_{\mathrm{osc}}}\right)^3~.
\ee
The existence of DM structure down to small scales requires that the axions behave as dark matter by the time the temperature of the Universe is $\sim \keV$, \ie\  $T_{\mathrm{osc}}\gtap 1\keV$.  Thus, there is an upper bound on the initial halo mass. More sophisticated analysis of the constraints on the axion isocurvature power spectrum at small scales can be found in Ref. \cite{Irsic:2019iff}.

Going forward we will assume that the axion makes up a sizable fraction of the dark matter abundance and place a bound on its self-coupling, equivalently $f_a$, through recurrent axinova.  There are four parameters that determine the amount of axion dark matter that is converted to dark radiation: the axion mass $m_a$, the axion self coupling $\lambda$ which in simple models is determined by the decay constant $f_a$, the structure mass $M_0$ (or equivalently $M_{peak}(z_{c})$), and the red-shift at which minihalos first form $z_c$.  Numerical simulations \cite{Vaquero:2018tib,Buschmann:2019icd} indicate that the white noise spectrum has large amplitude at small scales $A_0\sim \mathcal{O}(1)$ and thus minihalos form as early as possible $z_c\sim z_{\mathrm{eq}}$, with mass given by (\ref{eq:M0Tosc}).  

As times evolves, the characteristic mass grows as $M_h\sim (1+z)^{-2}$ as minihalos merge with each other.  Since a characteristic mass halo has concentration $c\approx 4$ its scale radius and density vary with redshift as $r_s\sim (1+z)^{-5/3}$, $\rho_s\sim (1+z)^3$ and consequently the speed at the scale radius depends on redshift as $v_s\sim (1+z)^{-1/6}$.  From Eqs.~(\ref{eq:taugrav}) and (\ref{eq:tauself}) this implies that the time scales for collapse scale as $\tau_{gr}\sim (1+z)^{-7}$, $\tau_{self}\sim (1+z)^{-19/3}$. This rapid lengthening of the axion star formation time as the Universe ages means that the dominant DM mass loss occurs as soon as the minihalo mass is larger than the critical star mass, and the earlier that occurs the greater the fraction lost.  More precisely, assuming $t_{\rm crit}$ is in the power law regime, the decay rate for halos of mass $M_0$ which initially form at redshift $z_c$ is,
\begin{equation}
\label{eq:dfdz}
\begin{split}
\frac{d f_{\rm decay}}{dz}&\sim
76500\pi^{2/3}\kappa\frac{M_{pl}^3 \overline{\rho}_{\rm col}^2}{M_0  f_a^5 m_a^4}\left(\frac{1+z}{1+z_c}\right)^8 \frac{1}{(1+z)^{5/2}H_0}
\\
&\times \left[1+
75\pi^{4/3}\left(\frac{f_a}{M_0^{1/3}\overline{\rho}_{\rm col}^{1/6}}\right)^4\left(\frac{1+z}{1+z_c}\right)^{2/3}\right]\\
&\times\left(\frac{\overline{\mstar}}{\mstar^{\rm max}}\right)^{\alpha-2}\Theta\left(M_{peak}(z)-\mstar^{\rm max}\right)~,
\end{split}
\end{equation}
where we have suppressed the logarithmic corrections to the Rutherford cross section in (\ref{eq:taugrav}), taken $b=d=1$, and $\overline{\rho}_{\rm col}=(1+z_c)^3 \overline{\rho}_0$ is the background density at the time of initial collapse. 


 The from of (\ref{eq:dfdz}) makes clear that the rate is peaked to early redshift and this rate is enhanced by decreasing both $m_a$ and $f_a$.  
If the timescale for scattering is set by self interactions, \ie\ $f_a \lsim M_0^{1/3} \overline{\rho}_0^{1/6}$, then along curves where $f_a\sim m_a^{-4/5}$ the decay rate is constant.  Furthermore, for any choice of parameters there is a maximal $f_a$ above which there is not enough time to form a critical mass star in a minihalo.
This leads to a region, bounded from below (above), in $m_a-f_a$ ($m_a-f_a^{-1}$) space which is constrained by the cosmological data discussed above (\ref{eq:decayfrac}).

\begin{figure}	
	\includegraphics[width=0.5\textwidth]{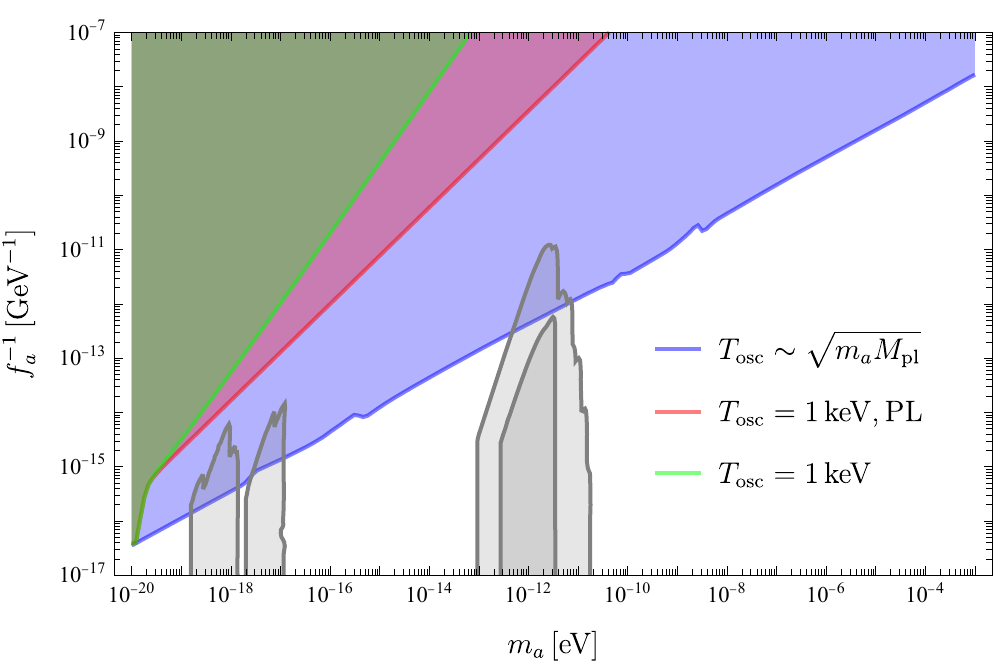}
	\caption{The exclusion region from axinovae for different assumptions for axion parameters, see text for more details.  Existing limits from the black hole superradiance are shown in grey. The green region is the most conservative bound using a constant rate of $d{\rm log}M/dt$ at $\mstar^{\rm max} \le \overline{\mstar}$ and a late $T_{\rm osc}$, with a formation timescale given in Eq.~(\ref{eq:tcrit})
The red region uses a power-law growth (PL) with $M\propto t^{2/3}$ over all the mass ranges and it also assumes the lowest oscillation temperature. The blue region presents the bound after optimizing over oscillation temperature.
	}
	\label{fig:axion_fa_alpha0.5}
\end{figure}



In Fig.~\ref{fig:axion_fa_alpha0.5}, we plot the region that is constrained by the axinovae, for various assumptions. The gray regions are excluded by black hole superradiance constraints \cite{Mehta:2020kwu, Baryakhtar:2020gao, Unal:2020jiy,AxionLimits}.
The most conservative (weakest) constraint, shown in green, comes from assuming that the oscillation temperature is low and that the time to reach a critical star is given by (\ref{eq:tcrit}).  Over most of the green region the critical star mass is low and the growth ($d\log M/dt$) is still in the constant regime.  Given constraints on large scale structure we take the lowest possible oscillation temperature to $\sim 1$ keV.  The later an axion starts oscillating the larger the mass of the initial axion miniclusters, which leads to a longer axion star production time $\tau$, suppressing the resulting appearance of axinova.  

In the red region we again assume the lowest possible oscillation but now assume that the star growth is power law, $M\sim \rho_s r_s^3 (t/\tau)^{2/3}$, for all star masses, as discussed in Appendix~\ref{app:powerlawinhalos}.  At masses below $\rho_s r_s^3$ the power law predicts faster growth than the constant growth assumed in (\ref{eq:tcrit}) and the green region.  This makes the bound  stronger.  For $f_a\gtap10^{15} \mathrm{GeV}$ the axion star critical mass is larger than where exponential growth transitions to power law in the green and the two constraints coincide. 

Finally, the blue region is the strongest constraint and is found by optimizing over the oscillation temperature.  The maximum possible oscillation temperature arises when the axion starts oscillating with its zero temperature mass, $T_{\rm osc} = \sqrt{m_a M_{\rm pl}}$.  These high temperatures will lead to the lightest axion miniclusters and the shortest star production times, but such miniclusters may not be massive enough to contain a critical star.  At each point in the parameter space, we select the highest possible $T_{\rm osc}$ that leads to a massive enough minicluster.  Since, $M_h \sim T_{\rm osc}^{-3}$ this selected temperature is still close to $\sqrt{m_a M_{\rm pl}}$.  In Fig.~\ref{fig:axion_fa_alpha0.2}, in Appendix~\ref{app:relic_abundance} we show the constraint for $\mstar\sim t^{1/5}$, when the leading order the decay rate is independent of $T_{\rm osc}$.

In the excluded regions an $\mathcal{O}(1)$ fraction of all dark matter has passed through an axionova.  This may lead to other observables in axion experiments or in cosmological observations. 
Given the high powers that appear in (\ref{eq:dfdz}) if the constraints on decaying dark matter are improved in the future the region of parameter space excluded will not be greatly altered.

\section{Conclusions}\label{sec:conclusions}
We obtain new bounds on axion dark matter parameters $m_a,f_a$ assuming the formation of dense axion minihalos, motivated by the post-inflationary scenario. Axion perturbations in the post-inflationary scenario will lead to the formation of dense substructures known as axion miniclusters or minihalos after matter-radiation equality, which can subsequently form coherent objects known as axion stars at the core of axion minihalos. 
Low mass dilute axion stars, supported by gradient pressure, can be cosmologically stable. However, they will accrete more axions from minihalos and continue to grow in mass until the axion self-coupling becomes important and the gradient pressure can no longer stop them from collapsing and emitting relativistic axions, in an \emph{axinova}.  
The remnant of an axinova is a less massive star which will again grow, leading to recurrent axinova.

If the recurrent formation rate is large enough and axinovae are active, they can convert a significant fraction of dark matter into radiation which can be constrained by measurements of large scale structure formation. 
Our constraint only depends on the axion self-coupling and gravity.  The self coupling can be mapped to axion-photon and axion-neutron couplings in specific models. Those constraints are obtained by requiring the population of dense axion stars formed in axion minihalos at high redshifts shall not dominate the mass of dark matter. If the axion is only a fraction of dark matter or only a few percent of axion dark matter is decaying, the conversion to dark radiation may be cosmologically significant in future observations but consistent with the current data.  Alternatively, if the axinova has a branching fraction into standard model states there may be observables in the region or parameter space close to our bound. We leave a more detailed study of the cosmological evolution or possible visible signals to future work.

\section*{Acknowledgements}
We thank Asimina Arvanitaki, Masha Baryakhtar, Nikita Blinov, Abhish Dev, George Fleming, Junwu Huang for helpful discussions.  We especially thank Josh Eby for comments on an early draft.  
PJF and HX are supported by Fermi Research Alliance, LLC under Contract DE-AC02-07CH11359 with the U.S. Department of Energy. N.W. is supported by NSF under award PHY-1915409, by the BSF under grant 2018140, and by the Simons Foundation.
HX thanks NYU for support and hospitality while a portion of this work was completed.
This work was performed in part at the Aspen Center for Physics, which is supported by the National Science Foundation grant PHY-1607611.

\onecolumngrid
\appendix
\section*{Appendix}
\section{Axion Star Configurations}\label{app:as_configuration}

The stable axion-field configuration for the gravitational bound-state of non-relativistic axions can be found by solving the Gross-Pitaevskii-Poisson equations, which must be done numerically.  For a thorough review, see \cite{Eby:2019ntd} and references therein.  However, it has been shown that a good approximation of these solutions is obtained by using a Gaussian ansatz for the field profile \cite{Chavanis:2011zi,Chavanis:2011zm,Chavanis:2016dab}.  Doing so gives some insight into the competing effects driving the physics \cite{Visinelli:2017ooc}.  Expanding the axion potential \eref{eq:VaxionFull} to quartic order one finds an attractive self interaction 
\be
V=\frac{1}{2}m_a^2 \phi^2 - \frac{\lambda}{4!}\phi^4~,
\ee
with $\lambda = (1-3c_{ud})m_a^2/f_a^2$.  An axion star of mass $\mstar$ and radius $\rstar$ has energy \be
E_* = -\frac{G_N \mstar^2}{\rstar} + c_1 \frac{\mstar}{2\,m_a^2 \rstar^2} - c_2 \frac{\lambda\mstar^2}{12\,m_a^4 \rstar^3}~.
\ee
In order, these terms correspond to the gravitational self energy, the gradient pressure, and the internal energy from self interactions.  The numerical coefficients, $c_i$, depend upon the details of the field profile and are found numerically \cite{Ruffini:1969qy,Membrado:1989ke,Visinelli:2017ooc} to be $c_1= 9.9$, $c_2=0.85$.  The mass-radius relation for axion stars, found by minimizing $E_*$, has two solutions 
\be
\label{eq:rstarpm}
\rstar^\pm=\frac{c_1}{2\,G_N \mstar m_a^2}\left(1\pm \sqrt{1-\frac{c_2}{c_1^2}  \lambda G_N\mstar^2}\right)~.
\ee
The $\rstar^+$ root corresponds to the so-called dilute branch and the axion field value is small.  On this branch gravitational attraction is balanced by gradient pressure leading to a stable configuration.  As is typical for objects supported by uncertainty pressure the product of the radius and mass of the star is a constant
\be
\label{eq:rstarp}
\rstar^+ = 9.9\frac{M_{\mathrm{pl}}^2}{m_a^2\mstar}~.
\ee
However, as one moves to larger axion star mass the self interactions cannot be ignored and if they are attractive (as asummed above) they destabilise the star.  There is a maximal mass, beyond which axion stars are no longer stable
\be\label{eq:mstarmax}
\mstar^{\mathrm max} = \frac{10.7}{\sqrt{\lambda}}M_{\mathrm{pl}}~.
\ee
The two solutions (\ref{eq:rstarpm}) meet at this maximal mass. The second solution is one where gravity can be ignored and the gradient pressure and the axion's attractive self interactions are in unstable equilibrium.  On this branch $\rstar\sim \mstar$.  

The value of the axion field at the center of the star scales as $a_0^2\sim \mstar/(m_a^2\rstar^3)$ so that at the low mass end of the $\rstar^-$ branch $a_0\sim 1$ and the axion field is not dilute.  The axions can no longer be thought of as non-relativistic and the solution is approximately constant density ($\rho\sim m_\pi^2 f_\pi^2$) and thus $\rstar\sim \mstar^{1/3}$.  However, it is believed that this field configuration is also unstable, with a lifetime \mbox{$\sim 10^3 m_a^{-1}$} \cite{Visinelli:2017ooc}, although alterations to the axion potential can make these solutions long lived \cite{Cyncynates:2021rtf,Kawasaki:2019czd,Olle:2020qqy}.

The upshot of this is that if a dilute axion star with mass below $\mstar^{\mathrm{max}}$ were to form and grow, by accumulation of additional axions, to the maximal mass it would then shrink in size and become a dense axion star which would survive for a short period.  During this time the dense axion star goes through several oscillations and a density singularity develops in the central core and this dense region emits relativistic axions lowering the density  \cite{Eby:2015hyx,Eby:2016cnq,Levkov:2016rkk}.  This process repeats and $\sim 30\%$ of the initial star mass can be emitted, leaving a dilute remnant which may in turn grow to the maximal mass and emit more relativistic axions.  Thus, maximal mass stars are an engine to turn substantial amounts of cold dark matter into radiation.

\section{The growth of axion stars in a large minihalo}\label{app:powerlawinhalos}
When the axion star mass larger than the characteristic star mass $\overline{\mstar}$, the mass growth is found to be well described by a power law, $\mstar \propto t^{1/\alpha}$. However, the growth rate at lighter masses in an axion minihalo is still unknown. One would expect the growth rate is larger at smaller radius in the minihalo environment due to the larger density and smaller velocity. If a star is formed within a small radius, the mass contained in this region is small. Therefore, lighter objects always start to form with a greater rate. For an NFW profile, the mass contained within $r$ is
\begin{equation}
	M(r)|_{r\rightarrow 0}=4\pi \rho_s r_s^3\left({\rm ln}\left(1+\frac{r}{r_s}\right)-\frac{r}{r+r_s}\right)\approx 2\pi \rho_s r_s r^2.
\end{equation}
The formation timescale given by self-interactions is $\tau_{self}\propto v^2/\rho^2$. At small radius of an NFW halo, the density and velocity scale as $\rho\propto 1/r$ and $v\propto \sqrt{r}$. Therefore, $M(t)\propto t^{2/3}$. Similarly, if the gravity dominates the axion star formation, $\tau_{gr}\propto v^6/\rho^2\propto r^5$ at small radius and we obtain the mass growth power law $M(t)\propto t^{2/5}$.  Since this scaling is active at short distance scales within the minihalo we consider a scenario where $\alpha = 3/2$ at all axion star masses, see Fig.~\ref{fig:axion_fa_alpha0.5}.


\section{Press-Schechter with White Noise-like Power at Short Distances}\label{app:PS_analysis}

We consider the density perturbations, $\delta\equiv \delta\rho/\overline{\rho}$, to consist of two contributions, conventional $\Lambda$CDM adiabatic perturbations that are present at all scales 
and isocurvature perturbations which are only become important over a finite range of scales. We take the isocurvature contribution to be a power low with a cut-off at very small scales, corresponding to a wavenumber $k_0$.  For the case of the axion it is believed the short-scale behavior has a power spectrum that is approximately that of white noise, corresponding to $n=3$ below.  Modes from these two contributions have different growth behaviors after they enter the horizon, in particular the adiabatic perturbations have logarithmic growth until matter-radiation equality while the isocurvature modes do not.  At late times, in the matter dominated era, they have similar growth.  Taking into account these different growth behaviors the two-point function of the density perturbations is
\be\label{eq:App-deltasq}
\langle\delta^2\rangle = \frac{2\pi^2}{k^3}\left(D_{adi}^2 I_1^2 L^2 A_s \left(\frac{k}{k_s}\right)^{n_s -1}+ D_{iso}^2 A_0 \left(\frac{k}{k_0}\right)^n\Theta(k_0-k)\right)~.
\ee
For a $\Lambda$CDM-like power spectrum $A_s\approx 2\times 10^{-9}$, $n_s\approx 0.97$, and the pivot scale is $k_s=5\times 10^{-3}$ Mpc$^{-1}$.  At late times $D_{adi}\approx D_{iso}\approx a/a_{eq} = (1+z_{\rm eq})/(1+z)$ and the exact forms can be found in standard references \eg\ \cite{Hu:1995en,Dodelson:2003ft}.  The constant $I_1\approx 9.1$ and $L\approx \log (0.1 a_{eq}/a)$.

The Press-Schechter formalism assumes spherical collapse of over-densities and that the probability for these collapses follows a Gaussian distribution whose variance, smoothed at some scale $R$, is given by
\be\label{eq:variance}
\sigma^2(z,R) = \int \frac{d^3k}{\left(2\pi\right)^3} \langle\delta^2\rangle \left|\widetilde{W}(kR)\right|^2~,
\ee
where $\widetilde{W}(kR)$ is the window function and can take various forms.  Here we focus on the so-called sharp $k$-filter where 
$\widetilde{W}(z)=\Theta(1-z)$.
For this choice of window function there is not a well defined mass, $M$, associated with the co-moving filter scale $R$, since the real space form of $\widetilde{W}$ does not have local support \cite{Maggiore:2009rv}.  However, we will follow the oft-used relation $M=6\pi^2 \overline{\rho}_0 R^3$ \cite{10.1093/mnras/262.3.627}, where $\overline{\rho}_0$ is the present day cosmological axion density.
Note that for (\ref{eq:variance}) to be well defined we have to introduce an IR cut-off $k_{IR}$ and we define $M_0 = 6\pi^2 \overline{\rho}_0 k_0^{-3}$.  
We are typically interested in halo masses and formation redshifts where the adiabatic perturbations are subdominant to the isocurvature perturbations, $A_s \ll A_0 $.  In this regime, once structures can form \ie\ $z<z_{\rm eq}$, the variance has the simple form
\be\label{eq:sigmasqMsimp}
\sigma^2(z,M) \sim \left(\frac{1+z_{\rm eq}}{1+z}\right)^2\frac{A_0}{n} \times \begin{cases}
	1 ~ & M  \le M_0 \\
	\left(\frac{M_0}{M}\right)^{n/3}~ & M >  M_0
\end{cases}~.
\ee

In the Press-Schechter approach the halo mass function is related to the probability to find $\delta>\delta_c\approx 1.686$, with the fraction of matter in objects of mass $M$ given by
\be\label{eq:dfdM}
\frac{df}{dM}= \sqrt{\frac{2}{\pi}}\frac{\delta_c}{M \sigma}\left|\frac{d\log\sigma}{d\log M}\right| e^{-\delta_c^2/\sigma^2}~.
\ee
The exponential suppression means that the most massive objects, with mass $M_{peak}$, to have formed are those for which $\sigma(z,M_{peak})= \delta_c$.  If the isocurvature perturbations were large enough, $A_0>n \delta_c$, these objects would form at $z_{\rm eq}$.  Instead, for more typical isocurvature perturbations of $A_0\approx 0.1$, the first halos to form are of mass $M_0$ and they form at 
\be\label{eq:zcollapse}
z_c \approx \sqrt{\frac{A_0}{n}}\frac{z_{\rm eq}}{\delta_c}~,
\ee
and subsequently grow, with the peak mass of the halo mass function being
\be\label{eq:Mpeak_app}
M_{peak}= M_0 \left(\frac{1+z_c}{1+z}\right)^{6/n}=M_0 \left(\frac{A_0}{\delta_c^2 n}\right)^{3/n}\left(\frac{1+z_{\rm eq}}{1+z}\right)^{6/n}~.
\ee

\section{Axion Relic Abundance from Misalignment}\label{app:relic_abundance}

We consider the relic abundance from the misalignment mechanism for an axion coupled to a dark confining gauge group ``DarkQCD", which is taken to be $SU(N_C)$ with $N_F$ vector-like quarks.  The temperature dependence of the mass is understood in two limits.  At low temperature the axion mass is independent of temperature and at high temperature the dilute instanton gas approximation is valid, leading to a power law dependence.  In between there could be a first or second order transition or a smooth cross over depending on $N_F, N_C$ \cite{Athenodorou:2022aay,Cui:2022vsr}.  For simplicity we take the temperature dependence mass to have the form
\be
m_a(T)=
\begin{cases}
	m_0 & T<T_c\\
	m_0 \left(\frac{T_c}{T}\right)^{b} & T\ge T_c
\end{cases}~.
\ee
Here we take the critical temperature to be the same as the confinement scale of DarkQCD, $T_c = \Lambda = \sqrt{m_0 f_a}$.  The dilute instanton gas approximation gives \mbox{$b=(11N_C+N_F-12)/6$}.  Taking $b$ large for temperatures in the vicinity of $T_c$ also approximates the form of a first order phase transition.
After PQ symmetry breaking, and before the instantons generate a potential for the axion, the misalignment angle $\theta=a/f$ has a flat potential and is free to take on any initial value in each causal patch.  The equation of motion for this angle is
\be\label{eq:EOMtheta1}
\ddot{\theta} + 3H\dot{\theta} + m^2(T) \theta =0~.
\ee
Assuming the cosmology is governed by a fluid with equation of state $p=\omega \rho$ (RD is $\omega =1/3$) then the scale factor $a\sim t^{\frac{2}{3(1+\omega)}}$ and $H=\frac{2}{3(1+\omega)t}$.  Combining this with the fact that temperature redshifts with the scale factor, $T\sim a^{-1}$, (\ref{eq:EOMtheta1}) becomes 
\be\label{eq:EOMtheta}
\ddot{\theta} + \frac{2}{(1+\omega)t}\dot{\theta} + m_0^2\left(\frac{\Lambda}{T_i}\right)^{2b}\left(\frac{t}{t_i}\right)^{\frac{4b}{3(1+\omega)}} \theta =0~.
\ee
This equation can be solved exactly by noting that $y=x^\alpha J_n(\beta x^\gamma)$ with $J_n$ the $n$-th Bessel function, satisfies the equation 
\be
\frac{d^2 y}{dx^2} - \frac{2\alpha-1}{x}\frac{dy}{dx} +\left(\beta^2\gamma^2x^{2(\gamma-1)}+\frac{\alpha^2-n^2\gamma^2}{x^2}\right) y =0~.
\ee
Thus, the solution to (\ref{eq:EOMtheta}) takes the form
\be
\left(\frac{t_i}{t}\right)^{\frac{1-\omega}{2(1+\omega)}}J_{\frac{3(\omega-1)}{2(2b+3(1+\omega))}}\left(m_0\, t\left(\frac{\Lambda}{T_i}\right)^b\frac{3(1+\omega)}{2b+3(1+\omega)}\left(\frac{t}{t_i}\right)^{\frac{2b}{3(1+\omega)}}\right)~.
\ee
Requiring that the argument of the Bessel function changes by an $\mathcal{O}(1)$ amount before oscillation is deemed to have set in, and identifying various powers of $t$ with $H$ and $m_a(T)$, the oscillation temperature is implicitly defined by
\be
m_{osc} \sim \frac{3+3\omega+2b}{2}H_{osc}~.
\ee
Notice that for large $b$, an axion mass that rapidly changes from zero to $m_0$ as can arise in a first order phase transition, this is different from the usual $m\sim 3H/2$ requirement since the rapid evolution of the axion mass provides its own ``friction".  From now on we consider the case of RD and thus $m_{osc} \sim (2+b)H_{osc}$.  We also consider the possibility that the dark sector and the SM are at different temperatures.  Assuming there are no thermalizing interactions between them, and ignoring the complication of different thresholds in the two sectors we take the ratio of temperatures to be a constant, $T_{D} = \xi T_{SM} \equiv \xi T$.  Thus, the oscillation temperature and mass are found by solving 
\be
\left(\frac{8\pi^3g_*(T)}{90}\right)^{1/2} \frac{T^2}{M_{pl}} = 
\begin{cases}
	\frac{2m_0}{3} &  \xi T < \Lambda \\
	\frac{m_0}{2+b} \left(\frac{\Lambda}{\xi T}\right)^b &  \xi T \ge \Lambda
\end{cases}~.
\ee  
If the oscillation begins while the mass is temperature dependent then 
\be
T_{osc} =\Lambda \left(c(T_{osc}) \frac{M_{pl}}{f_a}\frac{\xi^{-b}}{2+b}\right)^{\frac{1}{2+b}}\xrightarrow{b\rightarrow\infty}\frac{\Lambda}{\xi}~,
\ee
where $c(T) =  \sqrt{90/8\pi^3g_*(T)}$ and, assuming the SM dominates the energy density of the Universe, $0.06\lsim c(T) \lsim0.33$.
This solution is only consistent if $T_D > \Lambda$ which places the restriction $b\lsim \xi^2 M_{pl}/f_a$~\footnote{Note also that since confinement only occurs for a negative beta function $b\ge 5/3$.}.  For $b,\xi$ in violation of this bound the oscillation starts after the axion has attained its zero-temperature mass and $T_{osc}\sim \sqrt{2c(T_{osc}) m_0 M_{pl}/3}=\Lambda \sqrt{2c(T_{osc}) M_{pl}/3f_a}$. 

Once the oscillation temperature is known, and using the fact that ratio of axion number density to entropy density is constant,  the present day axion mass fraction can be determined:
\bea\label{eq:reliceq1}
\Omega_a &=& \frac{m_0 m_{osc}f_a^2\langle\theta^2\rangle}{2\rho_{crit}}\frac{g_*(T_0)T_0^3}{g_*(T_{osc})T_{osc}^3} = 
\sqrt{\frac{8\pi^3}{90g_*(T_{osc})}}\frac{m_0 f_a^2\langle\theta^2\rangle}{\rho_{crit}M_{pl}}\frac{g_*(T_0)T_0^3}{T_{osc}} 
\begin{cases} 
	\frac{3}{4} & \xi T_{osc}<\Lambda \\
	\frac{2+b}{2} & \xi T_{osc}\ge\Lambda
\end{cases}\\
&=&
\label{eq:reliceq2}
\frac{g_*(T_0)T_0^3}{\rho_{crit}M_{pl}} m_0^{1/2} f_a^{3/2} \langle\theta^2\rangle
\begin{cases}
	\frac{3}{4} \left(\frac{8\pi^3}{90}\right)^{3/4} \left(\frac{3f_a}{2M_{pl}}\right)^{1/2} g^{-1/4}_*(T_{osc}) & \xi T_{osc}<\Lambda \\
	\frac{2+b}{2} \left(\frac{8\pi^3}{90}\right)^{\frac{3+b}{2(2+b)}} \left(\frac{(2+b)\xi^b f_a}{M_{pl}}\right)^{\frac{1}{2+b}} g^{-\frac{b+1}{2(2+b)}}_*(T_{osc}) & \xi T_{osc}\ge\Lambda
\end{cases}~.
\eea

\begin{figure}
	\includegraphics[width=0.5\textwidth]{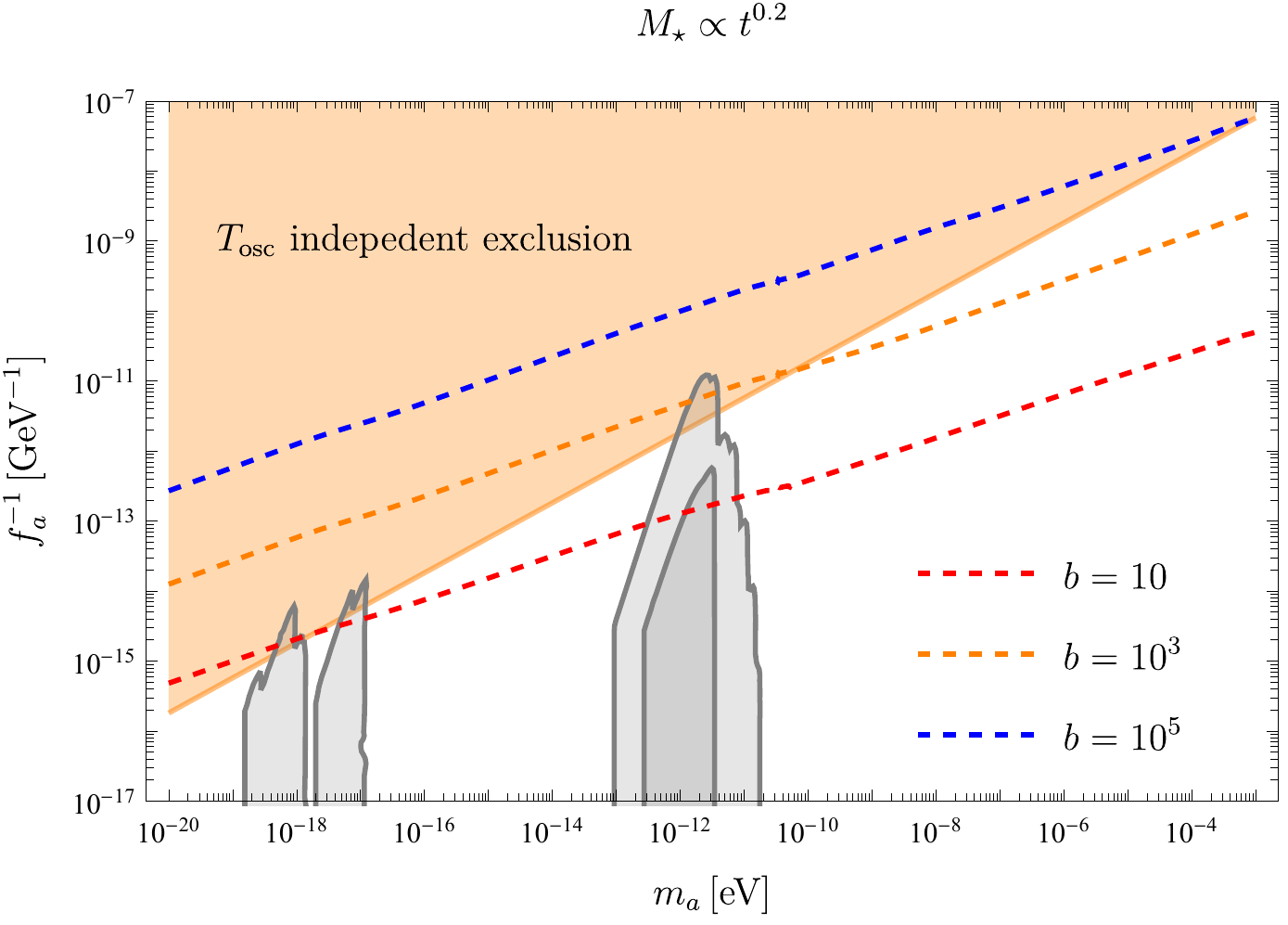}
	\caption{The exclusion plot of axion parameters from axinovae. The colored region represents the exclusion region assuming the axion star growth $M\propto t^{0.2}$, where the exclusion is independent of $T_{\rm osc}$. The gray regions are the existing limits from the black hole superradiance. Dashed curves are the axion parameters giving the correct relic abundance assuming axion mass behaves as $m_a(T)=m_a (\Lambda/T)^b$, where $\Lambda=\sqrt{m_af_a}$ is the dark confinement scale. A large $b$ can naturally come from a first-order dark QCD phase transition.
	}
	\label{fig:axion_fa_alpha0.2}	
\end{figure}

If the dark sector has roughly the same temperature as the standard model sector ($\xi\sim 1$), the confinement scale corresponds to a Hubble of $H\sim m_a f_a/M_{\rm pl}$, which is always smaller than $m_a$ because we require $f_a<M_{\rm pl}$ and the axion self-coupling is stronger than gravity. The axion mass will not be turned on until the dark confinement occurs. Therefore, $T_{\rm osc}$ is greatly delayed, which enhances the relic abundance since it is less diluted. The blue dashed curve in Fig.~\ref{fig:axion_fa_alpha0.2} shows the axion parameters that give the dark matter relic abundance assuming a slightly colder dark sector ($T_{\rm DS}= 0.5 T_{\rm SM}$) and axion mass to be turned on as $m_a\propto T^{-b}$. A large dark gauge group or a first-order phase transition in the dark sector will be needed for a large $b$. We also presented the $T_{\rm osc}$ independent constraint in Fig.~\ref{fig:axion_fa_alpha0.2} which assumes axion star mass grows like $M\propto t^{0.2}$, corresponding to $\alpha =5$.  For this value of $\alpha$ the decay rate (\ref{eq:dfdz}) is independent of $T_{\rm osc}$ in the region of parameter space dominated by self interactions.

While we have been focusing on a QCD-like axion model to study the relic abundance, there are other models that can enhance the self-coupling of axions while giving the correct relic abundance, such as a clockwork axion \cite{Kaplan:2015fuy} (discussed in Appendix \ref{app:enhanced_self}), friendship axion \cite{Cyncynates:2022wlq}, axions from dilute domain walls \cite{Redi:2022llj,Harigaya:2022pjd},and kinetic misalignment mechanism \cite{Co:2019jts,Chang:2019tvx}. In a clockwork axion scenario, a large field range is naturally produced for the axion field in the low-energy theory. The axion potential can have two confinement scales and two effective decay constants which can give the relic abundance that is needed while keeping the self-coupling strong. The friendship axion can resonantly convert the energy density in the axion sector with a larger decay constant to that with a lower decay constant if the mass ratio of two axions is close to 1. Therefore the relic density of axions with a low decay constant is greatly enhanced. Axion relic density can be greatly enhanced if the Peccei Quinn symmetry is followed by a period of inflation such that axion string networks are inflated away but will eventually reenter the horizon \cite{Redi:2022llj,Harigaya:2022pjd}. In this scenario, the decay of diluted domain walls occurs very late, enhancing the relic density of axions.
In kinetic misalignment \cite{Co:2019jts,Chang:2019tvx} the axion field does not start at rest but instead has a nonzero initial velocity.  The process of the axion settling into a minimum of the periodic potential, and generating an axion number density, is delayed since it can only occur after its initial kinetic energy has red-shifted away.  The initial velocity, $\dot{\theta}_i$, for the field is proportional to the net PQ charge and its generation requires an explicit breaking of the PQ symmetry at some scale.  This breaking should not be present at later times when the axion potential should be determined solely by instanton effects as can occur, for instance, if the breaking is from higher dimensional operators or arises from another scalar field acquiring a VEV.  The kinetic energy of the field becomes comparable to the potential energy when $\dot{\theta}_i (a_i/a)^3 \approx m_a(T)$, so large initial velocity and late generation both delay the onset of oscillations and increase the relic abundance.  
Kinetic misalignment tends to produce denser minihalos than conventional misalignment \cite{Eroncel:2022efc,Eroncel:2022vjg} due to a parametric resonance that enhances fragmentation \cite{Fonseca:2019ypl}.  If the fragmentation is not complete the power spectrum of axion density perturbations has features at many scales and our power law ansatz will not be a good approximation.  However, if the fragmentation completes before the kinetic motion is depleted the power spectrum is well approximated by white noise \cite{Eroncel:2022efc}.  In both cases the late-time halo mass function is peaked such that most of the mass is in mini-halos of mass $M_{peak}$. While there have been many models that can enhance either the axion relic abundance or the self-coupling, diluting the relic abundance is also possible in scenarios such as nonstandard thermal histories that lead to entropy production \cite{Nelson:2018via}. 

\section{Enhanced Axion Self-Coupling}\label{app:enhanced_self}

The axion self-coupling is given by $|\lambda| \sim m_a^2/f_a^2 \sim \Lambda^4/f_a^4$, assuming a cosine instanton potential. To obtain the right relic abundance for axion dark matter, $f_a$ is usually large since the relic abundance of axions is proportional to $f_a^2$. However, axion self-couplings can be enhanced without affecting the standard misalignment mechanism or the formation of axion miniclusters.
If the axion couples to two confining sectors, which can be naturally achieved with clockwork mechanism \cite{Kaplan:2015fuy}, the axion potential is 
\begin{equation}
V(a)=V_1(a)+V_2(a)=\Lambda_1^4\left(1-{\rm cos}\frac{a}{f_1}\right)+\Lambda_2^4\left(1-{\rm cos}\frac{a}{f_2}\right)~.
\end{equation}
Here $\Lambda_1, \Lambda_2$ are the confinement scales of the two strongly coupled sectors and $f_1,f_2$ are the corresponding decay constants. 

We consider the situation where the vacuum misalignment mechanism is mostly set by $V_1(a)$ and so we require $V_1'(a)\gg V_2'(a)$ and $V_1''(a)\gg V_2''(a)$ which corresponds to the requirements
\begin{equation}\label{eq:appcondition1}
\frac{\Lambda_1^4}{f_1}\gg \frac{\Lambda_2^4}{f_2}~,\,~\frac{\Lambda_1^4}{f_1^2}\gg \frac{\Lambda_2^4}{f_2^2}~.
\end{equation}
Satisfying these constraints will guarantee that the misalignment mechanism and the axion mass term and the rolling of axion field are solely determined by the strong sector with a confinement scale of $\Lambda_1$ and breaking scale $f_1$, which will be responsible for the relic abundance of the axion particles.
However, this does not fully determine the axion self-couplings. If $f_1\gg f_2$, the self-coupling can be dominated by the other strong sector, as long as the following condition is satisfied
\begin{equation}\label{eq:appcondition2}
\frac{\Lambda_1^4}{f_1^4}\ll \frac{\Lambda_2^4}{f_2^4}~.
\end{equation}
The conditions (\ref{eq:appcondition1}) and (\ref{eq:appcondition2}) can be consistent with each other provided $f_1\gg f_2$.  For instance, if $\Lambda_2/\Lambda_1 \equiv \epsilon \ll 1$ then $f_2/f_1 \sim \epsilon^\zeta$, with $1<\zeta<2$, will satisfy the conditions.
Assuming the strong coupling sectors satisfy these requirements then $m_a^2\sim\Lambda_1^4/f_1^2$ and $|\lambda|\sim\Lambda_2^4/f_2^4$ and the effective decay constant that labels the self-coupling strength is 
\begin{equation}
\tilde{f_a}=\frac{m_a}{\sqrt{|\lambda|}}=f_1\left(\frac{f_2 \Lambda_1}{f_1\Lambda_2}\right)^2\ll f_1.
\end{equation}
Therefore, the effective decay constant of an axion model that gives the self-coupling strength can be much smaller than the decay constant that is responsible for the relic abundance. They can be considered as two independent parameters.

\twocolumngrid
\bibliographystyle{apsrev4-2}
\bibliography{AxionStar}
\end{document}